# Dynamic, magnetic and electronic properties of the C14 Laves phase $Nb_{0.975}Fe_{2.025}$ compound: Mössbauer-effect study

Jan Żukrowski[1] and Stanisław M. Dubiel[2*]

[1]Academic Center for Materials and Nanotechnology, [2]AGH University of Science and Technology, Faculty of Physics and Applied Computer Science, al. A. Mickiewicza 30, 30-059 Krakow, Poland

**Abstract**

C14 Laves phase $Nb_{0.975}Fe_{2.025}$ compound was investigated by means of the Mössbauer spectroscopy. Spectra were recorded in the temperature range of 5-300K. Their analysis in terms of three sub spectra yielded information on magnetic and lattice dynamical properties of Fe atoms regularly occupying 2a and 6h lattice sites and, excessively, 4f sites. No indication of magnetism was observed down to the temperature of T≈50K, and spectral parameters viz. center shift, CS, and the main component of the electric field gradient, $V_{zz}$, behave regularly. In particular, analysis of CS(T) in terms of the Debye model yielded the following values of the Debye temperature, $T_D$: 453(5) K for the site 6h, 544(10) K for the site 2a, 479(4) K for the weighted average over 6h and 2a sites, and 363(35) K for the site 4f. Below ~50K anomalous behavior was observed: a broadening of the spectrum appeared indicating thereby a transition into a magnetic phase. Analysis of a temperature dependence of the hyperfine field, *B*, associated with the 6h and 2a sub spectra yielded the magnetic ordering temperature $T_{C1}$≈50K for the former and $T_{C2}$≈31K for the latter. The maximum values of the hyperfine field at 5K, $B_o$, were 8.2 kGs and 3.3 kGs, respectively. The $B_o$-values were used to estimate values the underlying magnetic moments, $\mu_{Fe}(6h)$=0.055-0.065 $\mu_B$ and $\mu_{Fe}(2a)$=0.02-0.025 $\mu_B$. Noteworthy, they are over one order of magnitude lower than those theoretically calculated. The CS and $V_{zz}$ parameters showed anomalies in the magnetic phase, in particular the former exhibited a strong departure from the Debye model prediction testifying to a significant effect of magnetism on the lattice vibrations.

PACS numbers: 63.20.dd, 63.20.kd, 75.50.Bb, 76.80.+y,

## 1. Introduction

The C14 hexagonal Laves $Nb_{1-x}Fe_{2+x}$ compounds belong to a family of Frank-Kasper phases. Despite their magnetic properties have been subject of several experimental and theoretical investigations since four decades a clear-cut picture concerning the magnetic phase diagram of the system has not been obtained. Main reasons for the situation are a high sensitivity of the magnetic properties to chemical composition in the vicinity of the stoichiometric one, on the one hand, and a high uncertainty of theoretical calculations concerning the magnetic structure and values of the magnetic moments (the latter are usually highly overestimated). Several experimental techniques were used to elucidate the magnetic properties of these Laves phase compounds including: nuclear magnetic resonance e. g. [1-2], Mössbauer spectroscopy e. g. [3-5], muon spin relaxation [5,6], DC and AC magnetic susceptibilities e. g. [4,7-10], Hall effect and transport properties [9,11] as well as the Compton scattering [12].

Crook and Cywinski proposed a magnetic phase diagram for the system in the compositional range $-0.03 \leq x \leq 0.04$, according to which the antiferromagnetic (AF) phase exists in the close vicinity of $x=0$ and the Néel temperature ranges between ~10 K ($x\approx-0.012$) and ~27K ($x\approx0.008$) [7]. The magnetic ground state is a mixture of AF and ferromagnetic (FM) phases. The most recent version of the magnetic phase diagram only partly agrees with the one just mentioned [9]. Namely, in both phase diagrams the FM ordering occurs for Fe- and Nb-over doped compounds, but border $x$-values and those of the Curie temperature, $T_C$, are different. Yet, the most significant difference exists in the vicinity of the stoichiometric composition i.e. $x=0$. Here, according to Crook and Cywinski, the AF and FM phases coexist, whereas in the phase diagram proposed by the authors of Ref. 9 the two FM phases are separated by a paramagnetic (PM) phase with a quantum critical point (QCP) located at $x\approx-0.015$, and a spin-density waves phase occurring between $-0.015 \leq x \leq 0.003$ at $T=0K$. In disagreement with both phase diagrams is a ferrimagnetic ordering recently suggested for an $Nb_{0.985}Fe_{2.015}$ compound based on a spin-dependent Compton scattering study supported by *ab initio* electronic structure calculations [12]. Noteworthy, some features characteristic of re-entrant spin-glasses (RSG) were observed both for the stoichiometric compound [10] as well as the one with a Fe-access [8].

Concerning theoretical studies some of them were concentrated on the QCP e. g. [13-14] whereas others depicted calculation of the magnetic structure and values of magnetic moments occupying various lattice sites e. g. [15-17]. Unfortunately, they have not helped to solve the enigma of the magnetic phase diagram of the NbFe2 system, because, in particular, the calculated values of the magnetic moments are, at least, one order of magnitude higher than the values estimated from experiment. However, precise experimental values are unknown as the neutron scattering experiment failed to deliver such information.

Motivated by the lack of the clear-cut nature and even controversies depicting the magnetism of the $Nb_{1-x}Fe_{2+x}$ compounds we performed Mössbauer spectroscopic study on an $Nb_{0.975}Fe_{2.025}$ sample. Such study can give site-resolved information on magnetic and lattice dynamic properties. The results we have obtained are presented and discussed in this paper.

## 2. Experimental

The alloy of the nominal composition $NbFe_2$ was produced from Fe (99.95 wt.%) and Nb (99.9 wt.%) components by levitation melting followed by subsequent casting into a pre-heated Cu crucible (15 mm in diameter) with a temperature of 1200°C. The temperature was kept at 1200°C for 45 min. and then the ingot was slowly cooled to room temperature with a rate of 5°C/min. Melting, casting and cooling was performed in an argon atmosphere. Impurity contents of non-metal elements C, N and O in the alloy was found to be less than 100 ppm. The phase identification and characterization of the sample obtained after the applied heat treatment was done by recording XRD patterns. The chemical composition of the final product was determined as $Nb_{0.975}Fe_{2.025}$ by EPMA measurements with a Joel JXA-8100 apparatus. Mössbauer spectra were measured in a transmission mode using a standard spectrometer (Wissel GmbH) and a drive operating in a sinusoidal mode. They were recorded in a 1024-channel analyzer in the temperature interval of 5-300K subdivided into two ranges: (a) 5-90K and (b) 80-300K. For the (a)-range a closed-cycle Janis refrigerator was used while for the (b)-range a standard Janis cryostat was employed. The temperature in both cases was kept constant to the accuracy better than ±0.1K. Examples of the measured spectra are shown in Fig. 1.

## 3. Results and discussion

### 3. 1. Spectra analysis

As can be seen in Fig. 1 the high-T spectrum has a shape of a quasi-symmetric doublet with a weak bulge on its higher velocity side, while the low-T spectrum is highly asymmetric and broadened. The latter feature can be associated with magnetism of the studied sample. In view of these features and taking into account the hexagonal structure of $NbFe_2$ the high-T spectra were analyzed assuming they are composed of two doublets corresponding to Fe atoms occupying two different lattice sites present in the unit cell of $NbFe_2$ viz. 6h and 2a with relative population 3:1. The latter was kept constant in the fitting procedure which was based on the transmission integral method. Additionally, a third sub spectrum (singlet) corresponding to the weak bulge was included into the fitting procedure. The relative amount of the third spectrum was fixed at the value of 2.5% found from the analysis of the 294K spectrum. This sub spectrum can be associated with Fe atoms occupying the 4f sites [3]. Free parameters were: line widths, Γ, spectral area of doublets (sextets), S, center shifts, CS, and the main components of the electric field gradient (having axial symmetry), $V_{zz}$ (for the doublets). This way of analysis worked well down to ~50K. At this temperature the quality of the fit, measured in terms of $\chi^2$, significantly deteriorated. It is worth mentioning here that the magnetic ordering temperature for the studied compound, as determined from magnetization measurements, was ~60 K [18]. Consequently, magnetic hyperfine interaction was included into the Hamiltonian i.e. the doublets were substituted by sextets. This step recovered the good quality of the fit for all spectra recorded at T ≤ ~50K. It should be added, that the axis of $V_{zz}$ was kept parallel to the hyperfine magnetic field, *B*.

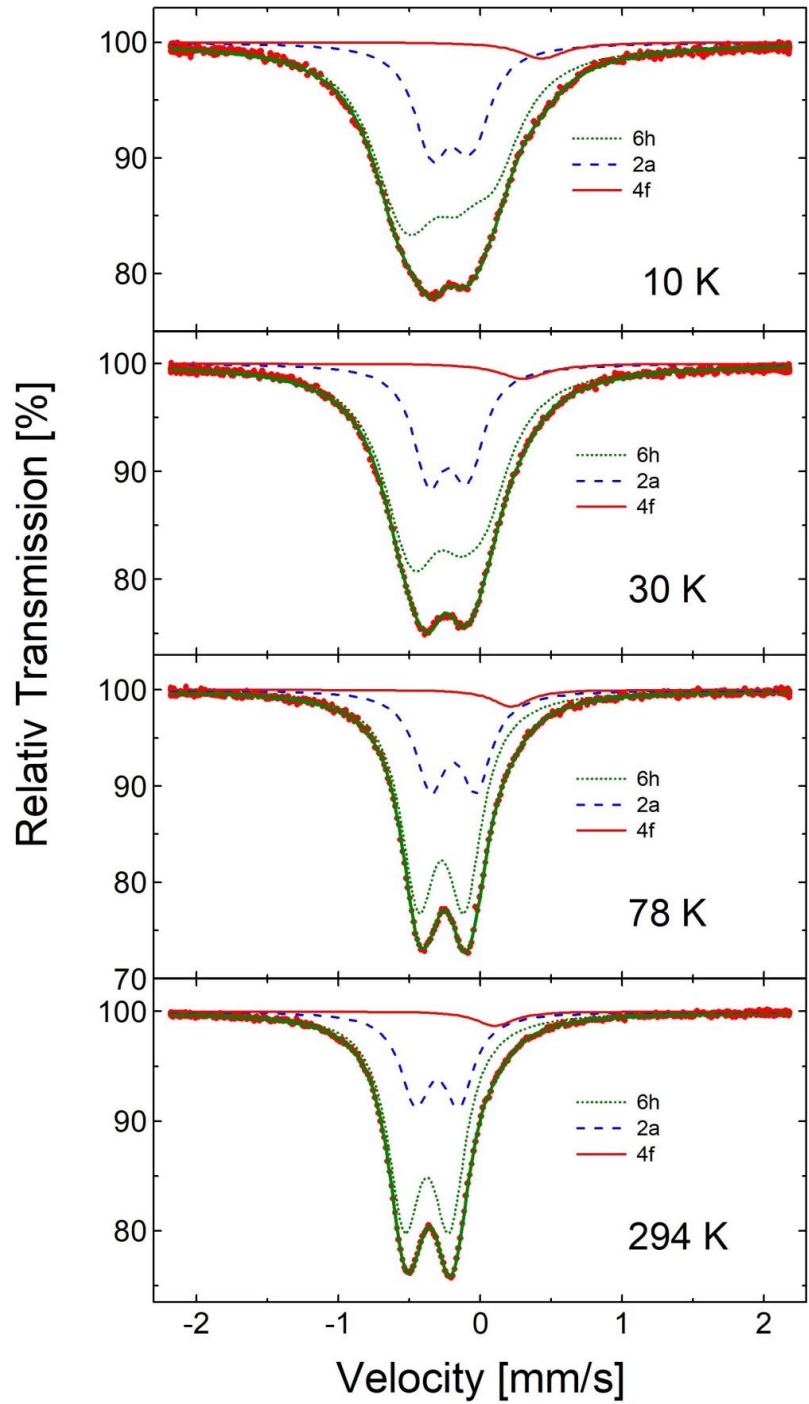

Fig. 1 Examples of the $^{57}$Fe Mössbauer spectra recorded at various temperatures shown. Three sub spectra associated with the 2a, 6h and 4f sites are indicated into which the spectra were decomposed.

Examples of the best-fit values of the spectral parameters for selected temperatures, as found with the applied fitting procedure, are displayed in Table 1.

**Table 1**

Best-fit spectral parameters for selected temperatures, $T$. CS are in mm/s (relative to a Co/Rh source kept at RT), $V_{xx}$ in $10^{18}$ V/cm$^2$, $B$ in kGs.

| T [K] | Site 6h | | | Site 2a | | | Site 4f |
|---|---|---|---|---|---|---|---|
| | CS | $V_{zz}$ | B | CS | $V_{zz}$ | B | CS |
| 10 | -0.271(2) | 0.291(2) | 7.5(3) | -0.217(4) | 0.158(3) | 3.5(3) | 0.41(2) |
| 30 | -0.267(1) | 0.242(2) | 5.6(2) | -0.225(2) | 0.163(3) | 1.2(7) | 0.30(1) |
| 51 | -0.256(2) | 0.207(4) | 2.6(2) | -0.230(5) | 0.169(9) | 0 | 0.221(8) |
| 78 | -0.272(2) | 0.194(1) | 0 | -0.195(3) | 0.186(3) | 0 | 0.217(8) |
| 150 | -0.300(1) | 0.192(1) | 0 | -0.211(2) | 0.185(2) | 0 | 0.171(7) |
| 294 | -0.375(1) | 0.186(1) | 0 | -0.295(2) | 0.187(2) | 0 | 0.103(6) |

Based on the CS-values and the spectral area lattice dynamical properties were studied while the values of the hyperfine fields, B, gave insight into the magnetic behavior of Nb$_{0.975}$Fe$_{2.025}$.

**3.2. Lattice dynamics**

Analysis of the Mössbauer spectra gives a unique opportunity for determining the Debye temperature, $T_D$, in two ways viz. from a temperature dependence of the center shift, CS(T), and from the recoilless factor, $f$. The $T$-dependence of CS can be expressed as follows:

$$CS(T) = IS + SOD(T) \qquad (1)$$

Where *IS* stands for the isomer shift and it hardly depends on $T$, and SOD(T) represents the so-called second-order Doppler shift. Its temperature dependence in the Debye model can be expressed by the following formula:

$$SOD(T) = \frac{3k_B T}{2mc}\left(\frac{3T_D}{8T} - 3\left(\frac{T}{T_D}\right)^3 \int_0^{T_D/T} \frac{x^3}{e^x - 1}dx\right) \qquad (2)$$

Where $m$ is the mass of a Fe atom, $k_B$ is the Boltzmann constant, $c$ is the speed of light, and $x = \hbar\omega/k_B T$ ($\omega$ being frequency of vibrations). The SOD-term is proportional to the mean-squared velocity of vibrations, <$v^2$>, hence to the kinetic energy of vibrating atoms. The analysis of the spectra we have performed enabled determination of $T_D$ for each of the three sites on which Fe atoms are present viz. 6h, 2a and 4f. Figure 2 shows the corresponding data for the sites 6h and 2a, in Fig. 3 is displayed the weighted average over 2a and 6h, <CS>$_{2a\_6h,}$ whereas Fig. 4 shows CS(T) obtained for the site 4f.

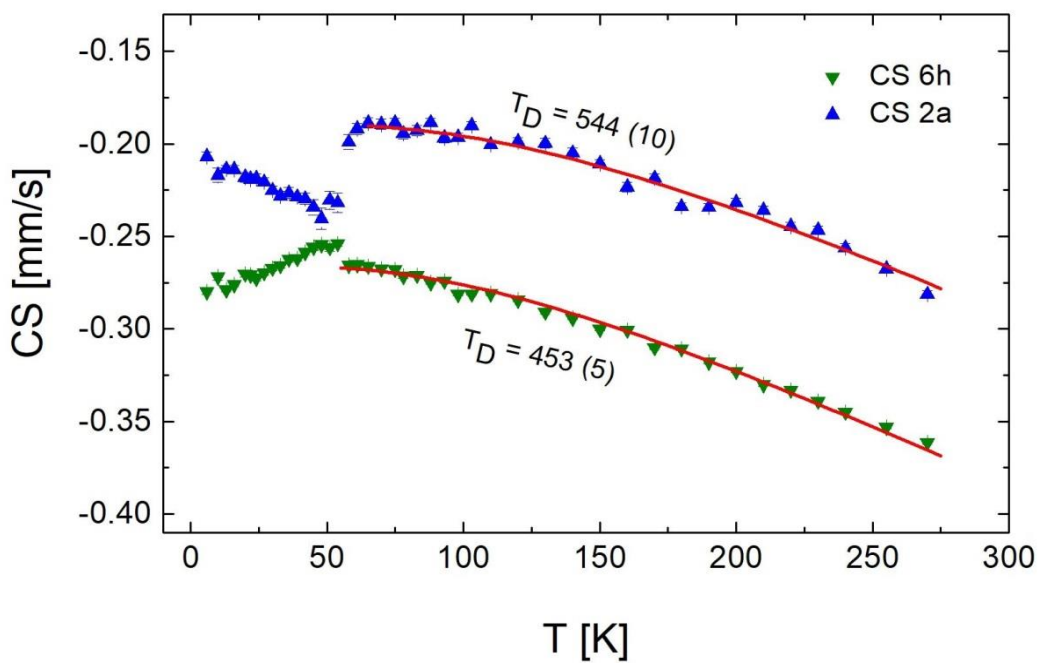

Fig. 2 Temperature dependence of the 2a and 6h sites center shifts, CS. The best-fit to the regular parts of the data is indicated by a solid line and the derived therefrom values of the Debye temperature are displayed.

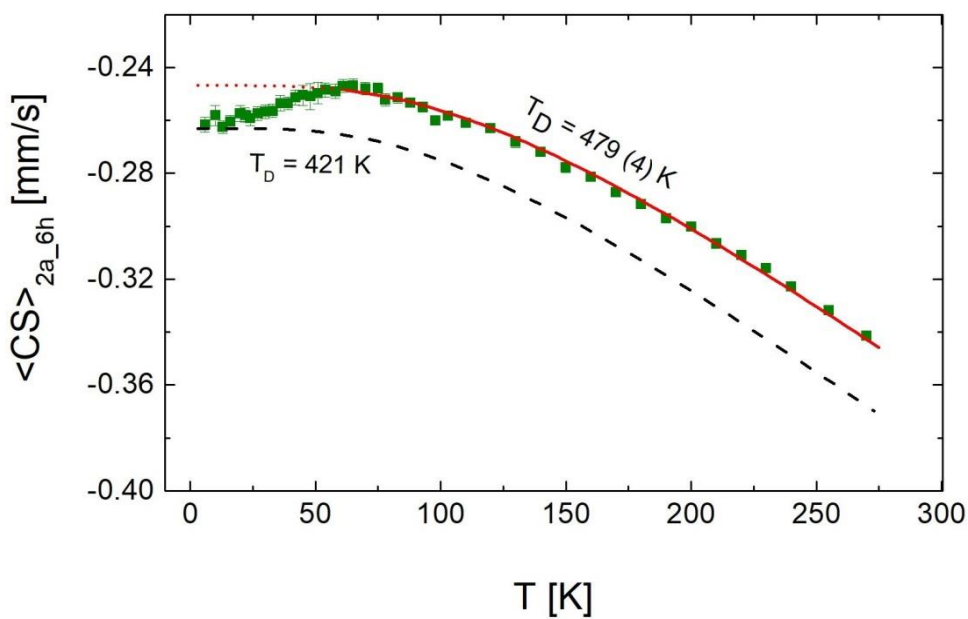

Fig. 3 Temperature dependence of the 2a and 6h site average center shift, $<CS>_{2a\_6h}$. The best-fit to the regular part of the data is indicated by a solid line, and the derived therefrom

Debye temperature, $T_D$=479(4) K. The behavior of the average CS corresponding to $T_D$=421 K is marked by a dashed line.

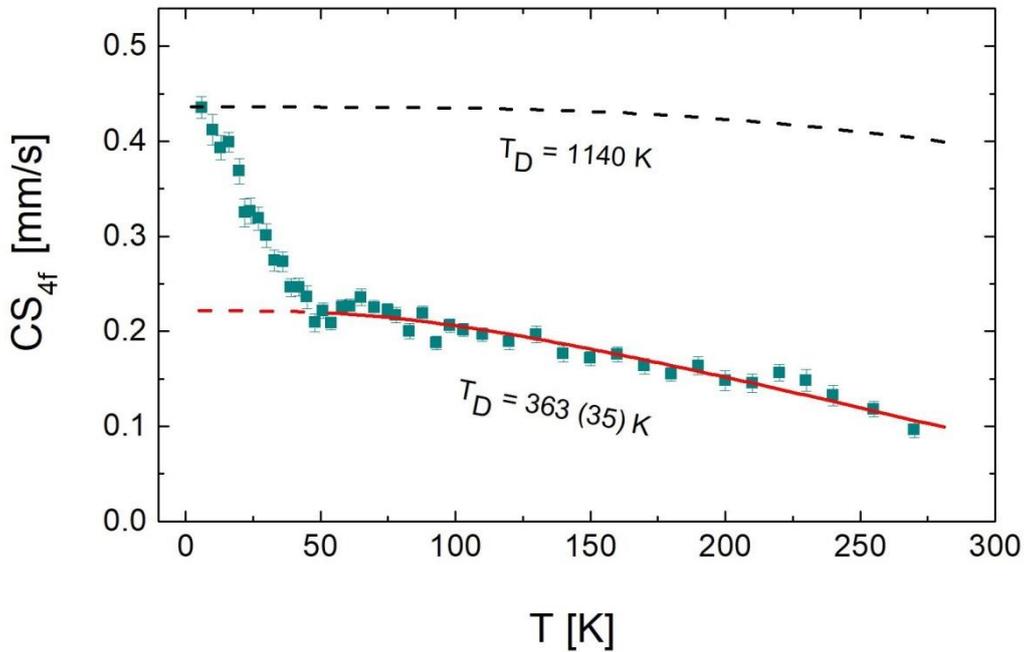

Fig. 4 Temperature dependence of the center shift of sub spectrum associated with Fe atoms on the 4f site, $CS_{4f}$. The best-fit to the regular part of the data is indicated by a solid line and the derived therefrom Debye temperature is displayed. The behavior of the $CS_{4f}$ corresponding to $T_D$=1140 K is marked by a dashed line.

It is evident that all sets of data exhibit a monotonic increase with the decrease of T down to ~50K, below which anomalies are observed. It should be here mentioned that our recent AC and DC magnetic susceptibilities measurements performed on the same sample yielded the value of ~60K for the magnetic ordering temperature [18]. This means that the anomalies displayed in Fig. 2 occur in the magnetic phase of the sample. The smooth parts of all four sets of CS-data were fitted to equation (2) yielding the following values for the Debye temperature: 453(5) K for the site 6h, 544(10) for the site 2a, 479(4) K for the weighted average over 6h and 2a sites, and 363(35) K for the site 4f. The obtained values of $T_D$ give clear evidence that the lattice dynamics of Fe atoms occupying different lattice sites in the $Nb_{0.975}Fe_{2.025}$ compound are significantly different. Noteworthy, the average value of $T_D$ is in line with recent theoretical calculations [17]. The anomalous behavior observed in the magnetic phase testifies to a magnetic origin of the revealed anomalies and is in line with Kim's theory according to which a spin-phonon interaction can be strongly enhanced in itinerant magnets [19] such the $NbFe_2$ [18]. Interestingly, CS(2a) significantly drops down on entering the magnetic phase to increase weakly with the decrease of T. On the contrary,

CS(6h) exhibits a weak increase near the magnetic ordering temperature followed by a gradual decrease on lowering *T*. The temperature behavior of <CS> - see Fig. 3, decreases monotonically below ~50K signifying a lattice softening. The difference between <CS(5K)> and <CS(60K)> is equal to ~0.02 mm/s which corresponds to an increase of $<v^2>$ by $1.2 \cdot 10^4$ $(m/s)^2$ or that of the kinetic energy by ~3.5 meV. The maximum softening of the lattice corresponds to a decrease of the Debye temperature by 58 K. Interestingly, the CS(4f), as shown in Fig. 4, exhibits an increase of ~0.22 mm/s which corresponds to a decrease of $<v^2>$ by $13.5 \cdot 10^4$ $(m/s)^2$ or that of the kinetic energy by 40 meV signifying lattice hardening. The latter can be expressed in terms of the Debye temperature viz. 1140K instead of 363K. The magnetic effect on the lattice dynamics can also be seen in Fig. 5. Here the temperature dependence of a spectral parameter, *f'*, proportional to the effective thickness, hence to the recoil-free fraction, *f*, is plotted. The regular part of the plot can be used to determine the Debye temperature which, within the Debye model approximation is related to *f* by the following formula:

$$\ln f' \propto -\frac{6E_R}{2k_B T_D}\left(\frac{1}{4}+\left(\frac{T}{T_D}\right)^2 \int_0^{T_D/T} \frac{x}{e^x-1}dx\right) \qquad (3)$$

Where $E_R$ is the recoil energy for the 14.4 keV gamma rays. The best-fit yielded $T_D$=385(5) K which is significantly less than the corresponding value determined from <CS>(T)-dependence. However, the result is not surprising as the Debye temperature derived from the *f*-factor is related to the mean-squared amplitude of vibrations whereas the one derived from the center shift is related to the mean-squared velocity of vibrations. Noteworthy, the $T_D$-values obtained from the *f*-factor are usually lower e. g. $T_D$(CS)=421 K vs. $T_D$(f)=358K for metallic iron [20] and $T_D$(CS)=403 K vs. $T_D$(f) =374 K for intermetallic σ-FeV [21].

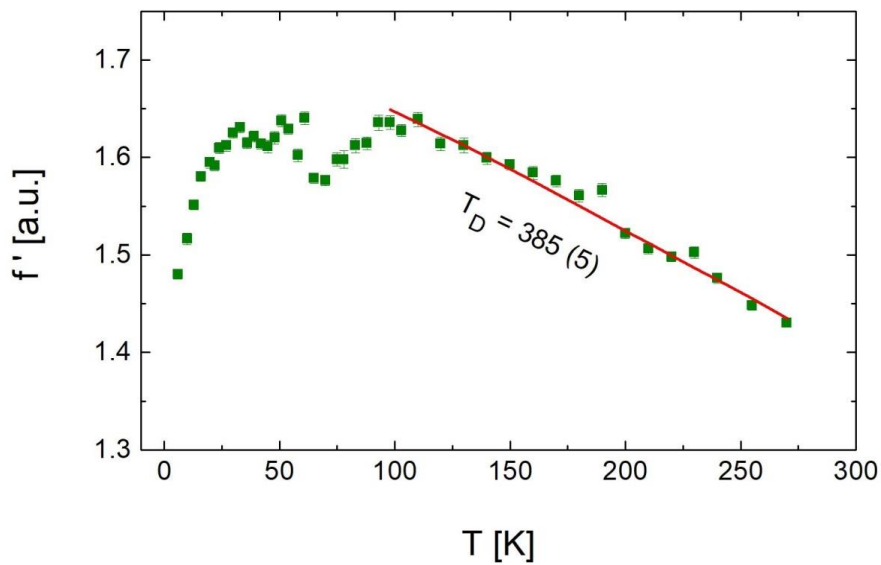

Fig. 5 The fitting parameter, *f'*, proportional to the recoil-free fraction, versus temperature for the two doublets associated with Fe atoms occupying the 2a and 6h sites. The solid line represents the best-fit to the regular segment of the data in terms of eq. (3).

### 3.3. Magnetic properties

The magnetic hyperfine field, *B*, has been regarded as a spectral parameter that is relevant to magnetic properties. Its temperature dependence can be readily used to determine a magnetic ordering temperature (Curie or Néel). From its magnitude a value of magnetic moments is often estimated, too. A well-defined sextet is characteristic of Mössbauer spectra measured on magnetically strong samples, while for weak magnets, like the present one, just a line broadening is observed in the magnetic state. If, in addition, a quadrupole interaction is present than extraction of the magnetic properties from Mössbauer spectrum is not an easy task. The procedure of the spectra analysis applied in the present case permitted determination of *B* for both lattice sites viz. 2a and 6h. In addition, by treating the linewidth of a singlet as a free parameter, we could find out whether or not non-zero spin-polarization exists on Fe atoms occupying the 4f sites. The temperature dependences of *B* obtained for the Fe atoms occupying the 2a and 6h sites are presented in Fig. 6. It is evident that below certain *T*-value, characteristic of the site, non-zero *B*-values appear with the maximum of *B(0)* ≈8 kGs for the 6h and *B(0)*≈3.5 kGs for the 2a sites, respectively. Assuming the hyperfine field is proportional to the spontaneous magnetization and following theories for itinerant magnets, the temperature dependence of *B* can be expressed by the following formulae [24]:

$$B(T) = B(0)\left[1 - \left(\frac{T}{T_C}\right)^{4/3}\right]^{1/2} \quad (4a)$$

$$B(T) = B(0)\left[1 - \left(\frac{T}{T_C}\right)^{2}\right]^{1/2} \quad (4b)$$

Where $T_C$ is a magnetic ordering temperature. The *B(T)* data for both sites could have been well-fitted to these equations.

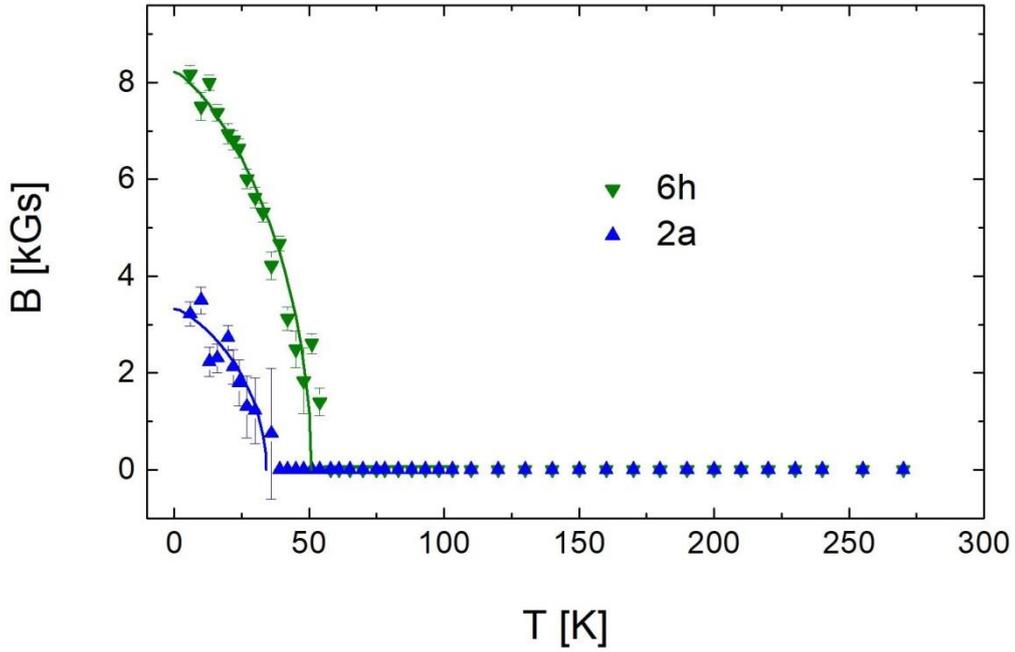

Fig. 6 The hyperfine magnetic field $B$, vs. temperature, $T$, as seen by Fe atoms on sites 2a and 6h. Solid lines represent the best-fits in terms of eqs. (4).

The best-fit parameters obtained by fitting the $B(T)$-data to eq. 4a are: $B(0)$=8.2 (7.6) kGs, $T_C$= 51 (50.2) K for the site 6h, and $B(0)$=3.3 (3.0) kGs, $T_C$= 33.7 (33.4) K for the site 2a. In brackets are given the corresponding values obtained with eq. 4b. Evidently, the two magnetic sub lattices have different magnetic ordering temperature, $T_C$, which is, to our best knowledge, a novel finding for the studied compound. The obtained $B(0)$-values can be next used to estimate a ratio of magnetic moments of Fe atoms residing on the two sites viz. R=$\mu_{Fe}$(2a)/$\mu_{Fe}$(6h). For that purpose one can assume, to the first approximation, that the hyperfine field, $B$, is proportional to the magnetic moment, $\mu$, i.e. $B = A \cdot \mu$. Assuming further the same value of the proportionality constant for both sites one gets R=B(2a)/B(6h)=0.4. This R-value can be subsequently used to validate various theoretical predictions concerning the magnetic structure of the ground state in the C14 NbFe$_2$ compound [15-17]. In particular, Subedi and Singh, based on first principles calculations, considered five co-linear spin configurations of Fe in this compound [15]. The energetically most favorable one, they found, was a ferromagnetic arrangement with the magnetic moments of $\mu_{Fe}$(2a)=1.18 $\mu_B$ and $\mu_{Fe}$(6h)=-0.75 $\mu_B$. However, the R-value for this structure is almost 3-fold higher than ours. On the other hand, the energetically least favorable was a ferromagnetic structure with $\mu_{Fe}$(2a)=0.57 $\mu_B$ and $\mu_{Fe}$(6h)=1.15 $\mu_B$ for which R=0.5, hence close to the value estimated here from B(2a)/B(6h) ratio. Remarkably, the ferromagnetic ordering was also considered for the Nb$_{0.985}$Fe$_{2.015}$ compound based on spin-dependent Compton scattering backed by *ab initio* calculations [12]. The calculated values of the magnetic moments were $\mu_{Fe}$(2a)=-0.997 $\mu_B$ and $\mu_{Fe}$(6h)=0.711 $\mu_B$ yielding R=1.4. It should be added that, as a rule, the values of the calculated magnetic moments for itinerant magnetic are significantly overestimated decreasing thereby their validity. Unfortunately, the neutron scattering technique has failed so far to deliver any meaningful data on the issue. Despite, the proportionality between the hyperfine field and the magnetic moment is not strictly obeyed and the proportionality constant is not universal [22] nevertheless it can be used to get an estimation on the values of $\mu_{Fe}$(2a) and $\mu_{Fe}$(6h). For this purpose we take for A the values obtained for σ-FeCr compounds which exhibit a similar type of magnetism as the presently studied compound i.e. weak and itinerant [23]. By doing so, one arrives at $\mu_{Fe}$(2a)=0.02-0.025 $\mu_B$ and $\mu_{Fe}$(6h)=0.055-0.065 $\mu_B$. For comparison, the ordered magnetic moment in the $Nb_{0.96}Fe_{2.04}$ compound was estimated as 0.02 $\mu_B$ [9], while for the Fe-under doped compounds as 0.035-0.078 per Fe atom [7]. In other words, the presently done estimation gave $\mu_{Fe}$-values which, on one hand, are in line with other experimental-based data, but in addition, we have obtained site-resolved values of the magnetic moments. Theoretical calculations predict that also Nb atoms occupying 4f sites have non-zero magnetic moments ranging between 0.16 and 0.48 $\mu_B$ [12,15-17]. These values are certainly overestimated. If correct, the corresponding values of the hyperfine field would lie between ~18 and ~50 kGs. However, a sub spectrum (singlet) associated with Fe atoms present on the 4f sites in the Fe-over doped compound does not split into sextet but merely broadens in the magnetic phase [3]. In the present case, as illustrated in Fig. 7, the linewidth in the paramagnetic state is 0.32 mm/s and the one at ~5K is 0.44 mm/s. This broadening corresponds to ~3 kGs or 0.04 $\mu_B$.

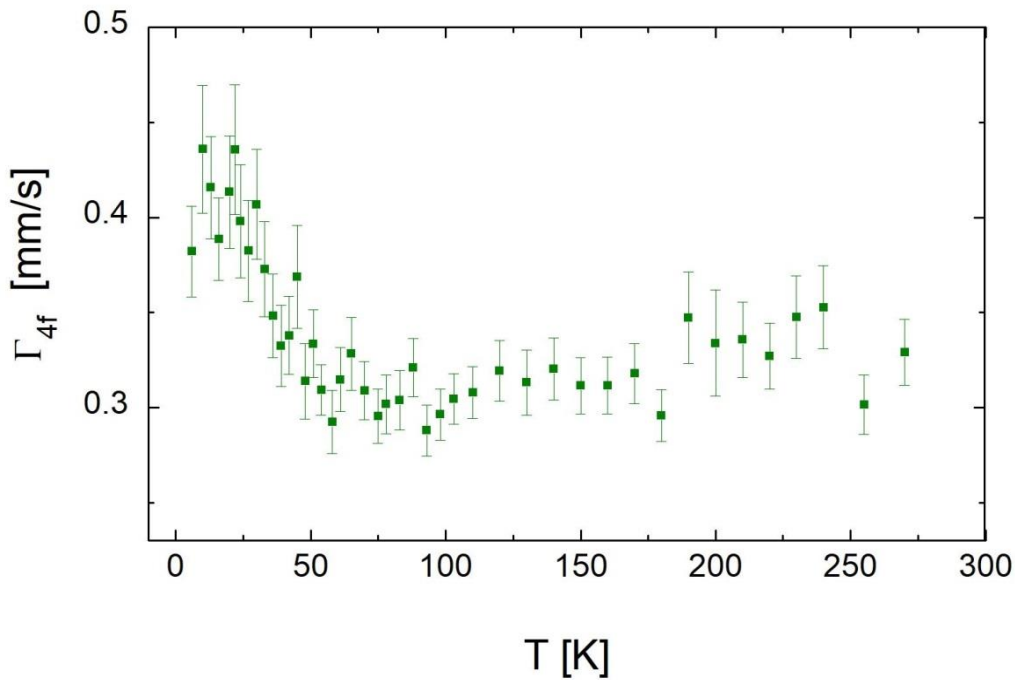

Fig. 7 Full linewidth at half maximum, $\Gamma_{4f}$, vs. temperature, T.

### 3.3. Electronic properties

### 3.3.1. Electric field gradient

The main component of the electric field gradient (EFG), $V_{zz}$, is plotted versus temperature in Fig. 8. In the paramagnetic phase there is no measurable difference in EFG for the 2a and 6h sites i.e. $V_{zz}(2a)=V_{zz}(6h)$. However, on transition into the magnetic phase anomalous behavior appears in both components, yet different. Namely, whereas the $V_{zz}(2a)$ shows a small step-like decrease at $T_C$ and remains constant down to 5K, the $V_{zz}(6h)$ monotonously increase below $T_C$ reaching ~50% increase at 5K.

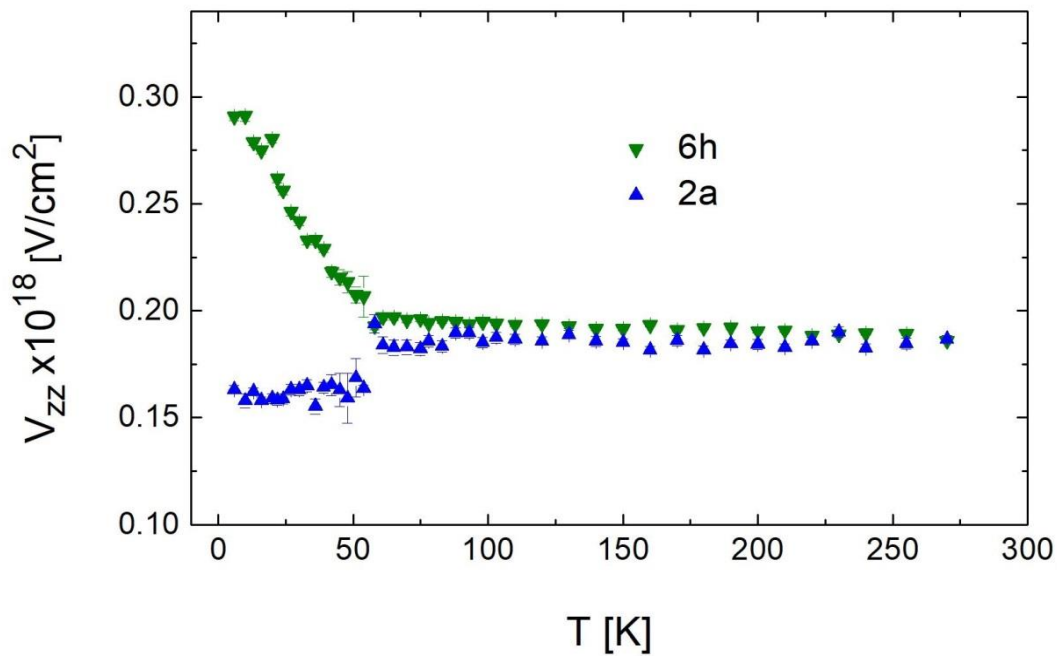

Fig. 8 The main component of the electric field gradient, $V_{zz}$, as a function of temperature, *T*, for the sites 2a and 6h.

### 3.3.2. Charge-density

One of the component of the center shift, CS, is the isomer shift, IS, which is related to a density of s-like electrons at Fe nuclei and it hardly depends on temperature. Consequently, any comparison of CS-values relevant to the charge-density must be done at constant T. Thus looking at the CS-values obtained from the analysis of the spectrum measured at 10K – see Table 1 – one can readily conclude that the charge-density is characteristic of the lattice site occupied by Fe atoms. The highest value of the charge have Fe atoms on the sites 6h, slightly lower those on the sites 2a, and very much lower the ones that reside on the 4f sites. One can even estimate charge-density differences between these sites using a scaling scheme proposed by Walker et al. according to which in metallic systems a change of 1s-like electron corresponds to a change of the isomer shift by ~2 mm/s [25]. Thus the charge-density on Fe atoms at 2a sites is by ~0.03 s-like electrons and that on Fe atoms at 4f sites by ~0.34 s-like electrons lower than the charge-density on Fe atoms at 6h sites. The latter signifies a charge flow from Fe atoms into Nb ones for which, according to calculations, the number of electrons is almost twice lower [26].

## 4. Conclusions

The measurements and their interpretation described in this paper permit to draw a general conclusion that lattice dynamical and magnetic properties as seen by probe Fe atoms are characteristic of a lattice site occupied by these atoms. Concerning the former ones:

1. In the paramagnetic phase, T≥~50 K, the lattice vibrations of all 3 lattice sites viz. 2a, 6h and 4f are in line with the Debye model.

2. The values of the Debye temperature, $T_D$, derived from a temperature dependence of the center shift, CS, are: 544(10) K for the site 2a, 453(5) K for the site 6h, 479(4) K for the weighted average over 6h and 2a sites, and 363(35) K for the site 4f.

3. In the magnetic phase, T ≤~50 K, strong anomalies in the CS(T) and in the electric field gradient behavior are observed. The former testify to a spin-phonon coupling.

4. Fe atoms present on all 3 lattice sites show weak, low temperature magnetism. The strongest one was revealed for 6h sites with $T_C$≈50 K and maximum hyperfine field $B_o$≈8 kegs, followed by that of 2a sites with $T_C$≈41 K and $B_o$≈3.5 kGs. Also excessive Fe atoms present on 4f sites exhibit sign of magnetism comparable to that seen on the 2a sites.

5. Fe charge-densities are also characteristic of the site: the highest charge-density was found for Fe atoms present on the 6h sites, slightly lower on the 2a sites and much lower on the sites 4f.

**Acknowledgement**

Dr. Frank Stein from the Max-Planck-Institut für Eisenforschung, Düsseldorf, is thanked for providing the sample for this study.